# On the invariant motions of rigid body rotation over the fixed point, via *Euler's* angles.

**Sergey V. Ershkov**

Institute for Time Nature Explorations,

M.V. Lomonosov's Moscow State University,

Leninskie gory, 1-12, Moscow 119991, Russia.

e-mail: sergej-ershkov@yandex.ru.

The generalized *Euler's* case (rigid body rotation over the fixed point) is discussed here: - the center of masses of non-symmetric rigid body is assumed to be located at the equatorial plane on axis *Oy* which is perpendicular to the main principal axis *Ox* of inertia at the fixed point. Such a case was presented in the rotating coordinate system, in a frame of reference fixed in the rotating body for the case of rotation over the fixed point (*at given initial conditions*).

In our derivation, we have represented the generalized *Euler's* case in the *fixed* Cartesian coordinate system; so, the motivation of our ansatz is to elegantly transform the proper components of the previously presented solution from one (rotating) coordinate system to another (fixed) Cartesian coordinates.

Besides, we have obtained an elegantly analytical case of general type of rotations; also, we have presented it in the *fixed* Cartesian coordinate system via *Euler's* angles.

**MSC classes:** 70E40 (Integrable cases of motion)

## 1. Introduction, equations of motion.

Euler's equations (dynamics of a rigid body rotation) are known to be one of the famous problems in classical mechanics, a lot of great scientists have been trying to solve such a problem during last 300 years.

Despite of the fact that initial system of ODE has a simple presentation, only a few exact solutions have been obtained until up to now [1-5], in a frame of reference fixed in the rotating body:

 - the case of symmetric rigid rotor {*two principal moments of inertia are equal to each other*} [1-3]: 1) *Lagrange's* case, or 2) *Kovalevskaya's* case;

- the *Euler's* case [4]: all the applied torques equal to zero (*torque-free precession of the rotation axis of rigid rotor*), the center of mass of rigid body coincides to the fixed point;

- the generalized *Euler's* case [5]: 1) the center of masses of non-symmetric rigid body is assumed to be located at the meridional plane which is perpendicular to the main principal axis *Ox* of inertia at the fixed point (*besides, the principal moments of inertia satisfy the simple algebraic equality*); 2) the center of masses of non-symmetric rigid body is assumed to be located at the equatorial plane on axis *Oy* which is perpendicular to the main principal axis *Ox* of inertia at the fixed point.

- other well-known but particular cases [6], where the existence of particular solutions depends on the choosing of the appropriate initial conditions.

The generalized *Euler's* case [5] was recently published, at the beginning of the year 2014. In our derivation, we should represent such a case in the *fixed* Cartesian coordinate system; so, the motivation of our ansatz is to elegantly transform the proper components of the previously presented solution from one (rotating) coordinate system to another (fixed) Cartesian coordinates.

Thus, finally we should answer how looks the motion in the *fixed* Cartesian coordinate system (?) if we obtain the appropriate expressions for the components of solution in a frame of reference fixed in the rotating body.

Also, we should note that the type 1) of the reported above [5] *generalized Euler's* case is proved to be one of the *particular* cases. Indeed, 2 important constants of such a solution (associated with 2 integrals of motions) are assumed to be mutually dependent one to each other. It means the proper restriction at choosing of one of the initial angular velocities at given initial positions of the rotating body in the fixed Cartesian coordinate system.

So, we should determine the appropriate structure of the solution in *Euler's* angles (which describe the proper motion in the fixed Cartesian coordinate system) only for the type 2) of the *generalized Euler's* case.

Let us remember the results of [5], concerning the type 2) of the *generalized Euler's* case. In accordance with [1-3], Euler's equations describe the rotation of a rigid body in a frame of reference fixed in the rotating body for the case of rotation over the fixed point as below (*at given initial conditions*):

$$\begin{cases} I_1 \dfrac{d\Omega_1}{dt} + (I_3 - I_2)\cdot\Omega_2\cdot\Omega_3 = P(\gamma_2 c - \gamma_3 b), \\ \\ I_2 \dfrac{d\Omega_2}{dt} + (I_1 - I_3)\cdot\Omega_3\cdot\Omega_1 = P(\gamma_3 a - \gamma_1 c), \qquad (1.1) \\ \\ I_3 \dfrac{d\Omega_3}{dt} + (I_2 - I_1)\cdot\Omega_1\cdot\Omega_2 = P(\gamma_1 b - \gamma_2 a), \end{cases}$$

- where $I_i \neq 0$ - are the principal moments of inertia (i = 1, 2, 3) and $\Omega_i$ are the components of the *angular velocity vector* along the proper principal axis; $\gamma_i$ are the components of the weight of mass *P* and *a, b, c* - are the appropriate coordinates of the center of masses in a frame of reference fixed in the rotating body (*in regard to the absolute system of coordinates X, Y, Z*).

Poinsot's equations for the components of the weight in a frame of reference fixed in the rotating body (*in regard to the absolute system of coordinates X, Y, Z*) should be presented as below [1-3]:

$$\begin{cases} \dfrac{d\gamma_1}{dt} = \Omega_3 \gamma_2 - \Omega_2 \gamma_3, \\[6pt] \dfrac{d\gamma_2}{dt} = \Omega_1 \gamma_3 - \Omega_3 \gamma_1, \\[6pt] \dfrac{d\gamma_3}{dt} = \Omega_2 \gamma_1 - \Omega_1 \gamma_2, \end{cases} \quad (1.2)$$

- besides, we should present the invariants of motion (*first integrals of motion*) as below

$$\begin{cases} \gamma_1^2 + \gamma_2^2 + \gamma_3^2 = 1, \\[6pt] I_1 \cdot \Omega_1 \cdot \gamma_1 + I_2 \cdot \Omega_2 \cdot \gamma_2 + I_3 \cdot \Omega_3 \cdot \gamma_3 = const = C_0, \\[6pt] \dfrac{1}{2}\left(I_1 \cdot \Omega_1^2 + I_2 \cdot \Omega_2^2 + I_3 \cdot \Omega_3^2\right) + P(a\gamma_1 + b\gamma_2 + c\gamma_3) = const = C_1. \end{cases} \quad (1.3)$$

So, system of equations (1.1)-(1.2) is proved to be equivalent to the system of equations (1.1), (1.3). It means that we could obtain the exact solutions of system (1.1), using the invariants (1.3).

## 2. Exact solution, a = c = 0 (*in a frame of reference fixed in the rotating body*).

Having chosen the additional invariant of motion (square of the vector of angular momentum) in [5], we supposed it to be valid for the system of equations (1.1)-(1.2) as below ($C_0 \neq 0$, $a = c = 0$, $I_1 \neq I_3$):

$$I_1^2 \cdot \Omega_1^2 + I_2^2 \cdot \Omega_2^2 + I_3^2 \cdot \Omega_3^2 = C_0^2$$

In such a case, we could obtain from the system of equations (1.1), (1.3):

$$\gamma_3 = \frac{I_3 \cdot \Omega_3}{C_0}, \qquad \gamma_2 = \frac{I_2 \cdot \Omega_2}{C_0}, \qquad \gamma_1 = \frac{I_1 \cdot \Omega_1}{C_0} \qquad (2.1)$$

- where [5]:

$$\Omega_3^2 = \left\{ \frac{C_2 \cdot C_3 + (I_2 - I_1) \cdot I_2 \cdot \Omega_2^2 - b \cdot I_2 \cdot C_3 \cdot \Omega_2}{(I_1 - I_3) \cdot I_3} \right\} \qquad (2.2)$$

$$C_2 = C_0 \cdot \left( \frac{C_1}{P} - \frac{C_0^2}{2P \cdot I_1} \right), \quad C_3 = \left( \frac{2P \cdot I_1}{C_0} \right),$$

- but the proper component of solution for $\Omega_2(t)$ in (2.2) is given by the re-inversed *quasi-periodic* function from the appropriate *elliptic* integral [7]:

$$\int \frac{d\Omega_2}{\sqrt{f_1(\Omega_2, \Omega_2^2)} \cdot \sqrt{f_2(\Omega_2, \Omega_2^2)}} = \int dt, \qquad (2.3)$$

- where ($I_1 \neq I_3$)

$$f_1 = \left( \left\{ \frac{2I_1 C_1 - C_0^2}{I_3 \cdot I_2} \right\} - \frac{2P \cdot I_1 \cdot b}{C_0 \cdot I_3} \cdot \Omega_2 + \frac{(I_2 - I_1) \cdot \Omega_2^2}{I_3} \right),$$

$$f_2 = \left( \left\{ \frac{C_0^2 - 2I_3 \cdot C_1}{I_1 \cdot I_2} \right\} + \frac{2P \cdot I_3 \cdot b}{C_0 \cdot I_1} \cdot \Omega_2 - \frac{(I_2 - I_3)}{I_1} \cdot \Omega_2^2 \right),$$

$$f_1(\Omega_2, \Omega_2^2) \cdot f_2(\Omega_2, \Omega_2^2) > 0 \; .$$

Besides, the appropriate component of solution for $\Omega_1(t)$ should be presented as below [5]:

$$\Omega_1^2 = \left\{ \frac{C_0^2 - 2I_3 \cdot C_1}{I_1 \cdot (I_1 - I_3)} \right\} + \frac{2P \cdot I_3 \cdot b \cdot I_2}{C_0 \cdot I_1 \cdot (I_1 - I_3)} \cdot \Omega_2 - \frac{I_2}{I_1} \cdot \frac{(I_2 - I_3)}{(I_1 - I_3)} \cdot \Omega_2^2 \qquad (2.4)$$

## 3. **Presentation of exact solution (a = c = 0), via *Euler's* angles.**

In accordance with [1-3], Euler's kinematic equations, which describe the rotation of a rigid body over the fixed point in regard to the fixed Cartesian coordinate system, should be presented as below (*at given initial conditions*):

$$\begin{cases} \Omega_1 = \dot{\psi} \cdot \gamma_1 + \dot{\theta} \cdot \cos\varphi, \\ \\ \Omega_2 = \dot{\psi} \cdot \gamma_2 - \dot{\theta} \cdot \sin\varphi, \\ \\ \Omega_3 = \dot{\psi} \cdot \gamma_3 + \dot{\varphi}, \end{cases} \qquad (3.1)$$

$$\begin{cases} \gamma_1 = \sin\theta \cdot \sin\varphi, \\ \gamma_2 = \sin\theta \cdot \cos\varphi, \\ \gamma_3 = \cos\theta \end{cases} \qquad (3.2)$$

- where ψ, θ, φ - are the appropriate angles, describing the positions of the reference fixed in the rotating body (*in regard to the absolute system of coordinates X, Y, Z*), see Fig.1:

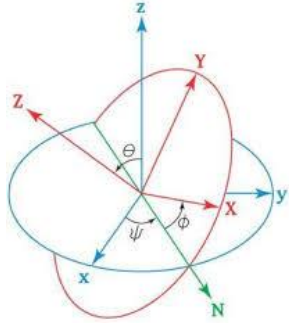

Fig. 1. Presentation of Euler's angles.

Equations (3.2) and (2.1) let us obtain as below:

$$\begin{cases} \varphi = \arctan(\gamma_1/\gamma_2), \\ \theta = \arccos\gamma_3 \end{cases} \Rightarrow \begin{cases} \varphi = \arctan\left(\dfrac{I_1\Omega_1}{I_2\Omega_2}\right), \\ \theta = \arccos\left(\dfrac{I_3\Omega_3}{C_0}\right), \end{cases} \qquad (3.3)$$

Then, the appropriate expression for the meaning of angle $\psi$ could be obtained from one of the Eqs. (3.1):

$$\dot{\psi} = \frac{\Omega_3 - \dot{\varphi}}{\gamma_3} \quad \Rightarrow \quad \dot{\psi} = \frac{C_0}{I_3}\cdot\left(1 - \frac{\dot{\varphi}}{\Omega_3}\right) \qquad (3.4)$$

Thus, formulae (3.3)-(3.4) are proved to describe the appropriate dynamics of rigid body rotation in regard to the absolute system of coordinates *X, Y, Z*, via *Euler's* angles.

## 4. **Analytical partial case of exact solution (a = c = 0), via *Euler's* angles.**

Let us assume that the proper simplifications are valid for expression (2.3) ($\alpha = const$):

$$I_1 = I_2, \quad C_0^2 - 2I_3 \cdot C_1 = 0, \quad I_1 > I_3, \quad \left\{ \frac{2I_1 C_1 - C_0^2}{I_1^2} \right\} = \alpha \frac{2P \cdot b}{C_0}, \quad \frac{2P \cdot b}{C_0} = \alpha \frac{(I_1 - I_3)}{I_3},$$

$$f_1 = \left( \left\{ \frac{2I_1 C_1 - C_0^2}{I_3 \cdot I_1} \right\} - \frac{2P \cdot I_1 \cdot b}{C_0 \cdot I_3} \cdot \Omega_2 \right), \quad f_2 = \left( \frac{2P \cdot I_3 \cdot b}{C_0 \cdot I_1} \cdot \Omega_2 - \frac{(I_1 - I_3)}{I_1} \cdot \Omega_2^2 \right), \qquad (4.1)$$

$$f_1(\Omega_2, \Omega_2^2) \cdot f_2(\Omega_2, \Omega_2^2) = \frac{I_1}{I_3} \cdot \left( \left\{ \frac{2I_1 C_1 - C_0^2}{I_1^2} \right\} - \frac{2P \cdot b}{C_0} \cdot \Omega_2 \right) \cdot \frac{I_3}{I_1} \cdot \left( \frac{2P \cdot b}{C_0} - \frac{(I_1 - I_3)}{I_3} \cdot \Omega_2 \right) \cdot \Omega_2 =$$

$$= \left( \frac{2P \cdot b}{C_0} \cdot \Omega_2 - \left\{ \frac{2I_1 C_1 - C_0^2}{I_1^2} \right\} \right) \cdot \left( \frac{(I_1 - I_3)}{I_3} \cdot \Omega_2 - \frac{2P \cdot b}{C_0} \right) \cdot \Omega_2 > 0 \; .$$

So, we should consider one of *particular* types of solutions of the *generalized Euler's* case [5] for the symmetric rotating rigid body ($I_1 = I_2$). It means the proper restriction at choosing of (one of) the initial angular velocities at given initial positions of the rotating body in the fixed Cartesian coordinate system.

Besides, in such a case the proper component of solution for $\Omega_2(t)$ in (4.1) could be obtained according to (2.3) as below:

$$\int \frac{d\Omega_2}{\sqrt{f_1(\Omega_2, \Omega_2^2)} \cdot \sqrt{f_2(\Omega_2, \Omega_2^2)}} = \int dt \; ,$$

- where ($I_1 > I_3$)

$$C_0^2 = 2I_3 \cdot C_1, \quad \alpha = \sqrt{2C_1 \cdot \left(\frac{I_3}{I_1^2}\right)}, \quad b = \left(\frac{C_1}{P}\right) \cdot \frac{(I_1 - I_3)}{I_1}, \quad \frac{2Pb}{C_0} = \left(\frac{2C_1}{\sqrt{2I_3 \cdot C_1}}\right) \cdot \frac{(I_1 - I_3)}{I_1},$$

$$\int \frac{d\Omega_2}{\sqrt{\frac{2P \cdot b}{C_0} \cdot (\Omega_2 - \alpha) \cdot \frac{(I_1 - I_3)}{I_3} \cdot (\Omega_2 - \alpha) \cdot \Omega_2}} = \int dt, \Rightarrow \left(A = \sqrt{\left(\sqrt{\frac{2C_1}{I_3}}\right) \cdot \frac{(I_1 - I_3)^2}{I_1 \cdot I_3}}, \quad u = \sqrt{\Omega_2}\right),$$

$$\Rightarrow 2\int \frac{du}{\alpha - u^2} = -A \cdot \int dt, \quad (0 \le u < \sqrt{\alpha}) \Rightarrow \left(\frac{2}{2\sqrt{\alpha}}\right) \cdot \ln\left|\frac{\sqrt{\alpha} + u}{\sqrt{\alpha} - u}\right| = -A \cdot t, \Rightarrow$$

$$\frac{\sqrt{\alpha} + u}{\sqrt{\alpha} - u} = \exp\left(-(A\sqrt{\alpha}) \cdot t\right), \quad u(t) = \left(\frac{\exp\left(-(A\sqrt{\alpha}) \cdot t\right) - 1}{\exp\left(-(A\sqrt{\alpha}) \cdot t\right) + 1}\right) \cdot \sqrt{\alpha} \Rightarrow \Omega_2(t) = u^2 \quad (4.2)$$

Using (4.2), we could obtain from (2.4) the appropriate expression for $\Omega_1$ ($I_1 = I_2$):

$$\Omega_1 = \sqrt{\left(\frac{\sqrt{2 \cdot I_3 \cdot C_1}}{I_1}\right) \cdot \Omega_2 - \Omega_2^2}, \quad \Omega_2 < \sqrt{2C_1 \cdot \left(\frac{I_3}{I_1^2}\right)} \quad (4.3)$$

- but, in addition to this, Eqs. (2.2) and (4.1)-(4.2) yield ($I_1 = I_2$):

$$\Omega_3 = \sqrt{\frac{2C_1}{I_3} \cdot \left(1 - \frac{I_1}{\sqrt{2I_3 \cdot C_1}} \cdot \Omega_2\right)}, \quad \Omega_2 < \frac{\sqrt{2I_3 \cdot C_1}}{I_1}, \quad \Omega_2 = u^2 \quad (4.4)$$

Thus, we have obtained the analytical expressions for all the components of angular velocities (4.2)-(4.4) in case $I_1 = I_2$, also we could obtain from (3.3) the appropriate expressions for the *Euler's* angles φ, θ

$$\begin{cases} \varphi = \arctan\left(\sqrt{\left(\dfrac{\sqrt{2 \cdot I_3 \cdot C_1}}{I_1}\right) \cdot \left(\dfrac{1}{u^2}\right) - 1}\right), & u(t) = \sqrt{\Omega_2} \\ \\ \theta = \arccos\left(\sqrt{1 - \dfrac{I_1}{\sqrt{2 I_3 \cdot C_1}} \cdot u^2}\right), & u(t) = \sqrt{\Omega_2} \end{cases} \qquad (4.5)$$

As for the dynamics of *Euler's* angle $\psi$, we should solve the ordinary differential equation of the 1-st order [8] as below, according to (3.4):

$$\dot{\psi} = \dfrac{\sqrt{2 I_3 \cdot C_1}}{I_3} \cdot \left(1 - \dfrac{\dot{\varphi}}{\Omega_3}\right) \Rightarrow \dot{\psi} = \sqrt{\dfrac{I_1}{\sqrt{2 \cdot I_3 \cdot C_1}}} \cdot \left(\sqrt{\left(\dfrac{\sqrt{2 \cdot I_3 \cdot C_1}}{I_1}\right) \cdot \dfrac{2 C_1}{I_3}} + \dfrac{\dot{\Omega}_2}{2\sqrt{\Omega_2} \cdot \left(1 - \dfrac{I_1}{\sqrt{2 I_3 \cdot C_1}} \cdot \Omega_2\right)}\right)$$

$$\Rightarrow \int d\psi = \sqrt{\dfrac{I_1}{\sqrt{2 \cdot I_3 \cdot C_1}}} \cdot \left(\sqrt{\left(\dfrac{\sqrt{2 \cdot I_3 \cdot C_1}}{I_1}\right) \cdot \dfrac{2 C_1}{I_3}} \cdot t + \int \dfrac{d\Omega_2}{2\sqrt{\Omega_2} \cdot \left(1 - \dfrac{I_1}{\sqrt{2 I_3 \cdot C_1}} \cdot \Omega_2\right)}\right)$$

$$\psi = \sqrt{\dfrac{I_1}{\sqrt{2 \cdot I_3 \cdot C_1}}} \cdot \left(\sqrt{\left(\dfrac{\sqrt{2 \cdot I_3 \cdot C_1}}{I_1}\right) \cdot \dfrac{2 C_1}{I_3}} \cdot t + \int \dfrac{du}{1 - \dfrac{I_1}{\sqrt{2 I_3 \cdot C_1}} \cdot u^2}\right) \quad (0 \le u < \sqrt{\dfrac{\sqrt{2 \cdot I_3 \cdot C_1}}{I_1}}) \Rightarrow$$

$$\psi = \left(\dfrac{I_1}{I_3}\right) \alpha \cdot t + \dfrac{1}{2} \ln\left(\dfrac{(\sqrt{\alpha}) + u}{(\sqrt{\alpha}) - u}\right), \quad \alpha = \sqrt{2 C_1 \cdot \left(\dfrac{I_3}{I_1^2}\right)}, \quad u(t) = \sqrt{\Omega_2} \qquad (4.6)$$

## 5. Discussion.

We discuss here the generalized *Euler's* case [5], which was recently published: the center of masses of non-symmetric rigid body is assumed to be located at the equatorial plane on axis *Oy* which is perpendicular to the main principal axis *Ox* of inertia at the fixed point. Such a case was presented [5] in the rotating coordinate system, in a frame of reference fixed in the rotating body for the case of rotation over the fixed point (*at given initial conditions*).

In our derivation, we have represented the generalized *Euler's* case [5] (2.2)-(2.4) in the fixed Cartesian coordinate system (3.3)-(3.4); so, the motivation of our ansatz is to elegantly transform the proper components of the previously presented solution from one (rotating) coordinate system to another (fixed) Cartesian coordinates.

Besides, we have obtained an elegantly analytical case (4.2)-(4.4) of general type of solutions (rotations); also, we have presented it in the fixed Cartesian coordinate system (4.5)-(4.6) via *Euler's* angles.

Thus, finally we should answer how the motion looks in the *fixed* Cartesian coordinate system if we obtain the appropriate expressions for the components of solution in a frame of reference fixed in the rotating body (see the next section).

Also, we should note that the case above of the generalized *Euler's* solution [5] is assumed to be one of the *particular* cases. Indeed, two of important constants of such a solution (*associated with 2 integrals of motions*) are assumed to be mutually dependent one to each other. It means the proper restriction at choosing of one of the initial angular velocities at given initial positions of the rotating body in the fixed Cartesian coordinate system.

## 6. Conclusion, final presentation of solution.

We have obtained absolutely new presentation (4.2)-(4.4) of exact solutions of the generalized *Euler's* case [5], which have been presented in the fixed Cartesian coordinate system (4.5)-(4.6) via *Euler's* angles.

We schematically imagine at Figs. 2-8 the dynamics of the components of solution (4.5)-(4.6) as presented below:

$$\begin{cases} \varphi = \arctan\left(\sqrt{\left(\dfrac{\alpha}{u^2}\right) - 1}\right), & \alpha = \sqrt{2C_1 \cdot \left(\dfrac{I_3}{I_1^2}\right)} \\ \theta = \arccos\left(\sqrt{1 - \dfrac{u^2}{\alpha}}\right), & u(t) = \sqrt{\Omega_2} \end{cases} \quad (6.1)$$

$$u(t) = \left(\dfrac{\exp\left(-(A\sqrt{\alpha})\cdot t\right) - 1}{\exp\left(-(A\sqrt{\alpha})\cdot t\right) + 1}\right) \cdot \sqrt{\alpha}, \quad A = \sqrt{\alpha \cdot \dfrac{(I_1 - I_3)^2}{I_3^2}},$$

$$\psi = \left(\dfrac{I_1}{I_3}\right)\alpha \cdot t + \dfrac{1}{2}\ln\left(\dfrac{\left(\sqrt{\alpha}\right) + u}{\left(\sqrt{\alpha}\right) - u}\right), \quad \alpha = \sqrt{2C_1 \cdot \left(\dfrac{I_3}{I_1^2}\right)}, \quad u(t) = \sqrt{\Omega_2} \quad (6.2)$$

Let us choose in Eqs. (6.1)-(6.2): $\alpha = 1$, just for simplicity of presentations.

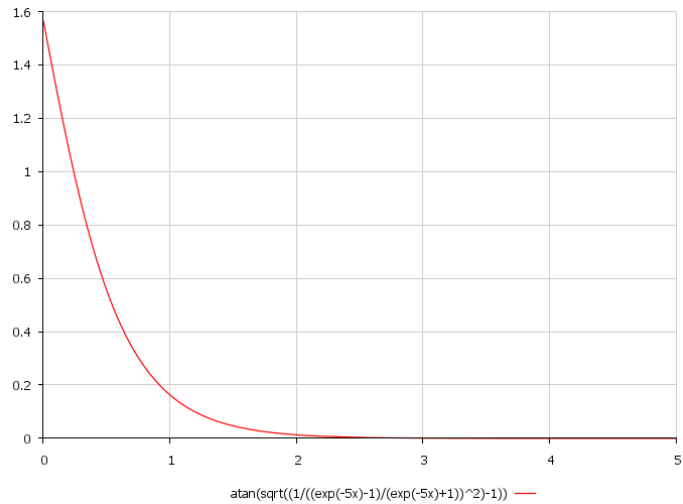

Fig.2. Function $\varphi(t)$: $A = 5$, $\alpha = 1$, see Eqs. (6.1).

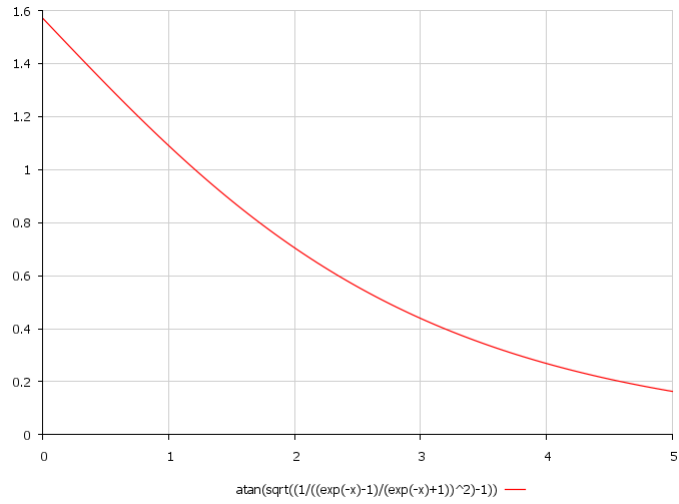

Fig.3. Function φ(*t*): $A = 1$, $\alpha = 1$, see Eqs. (6.1).

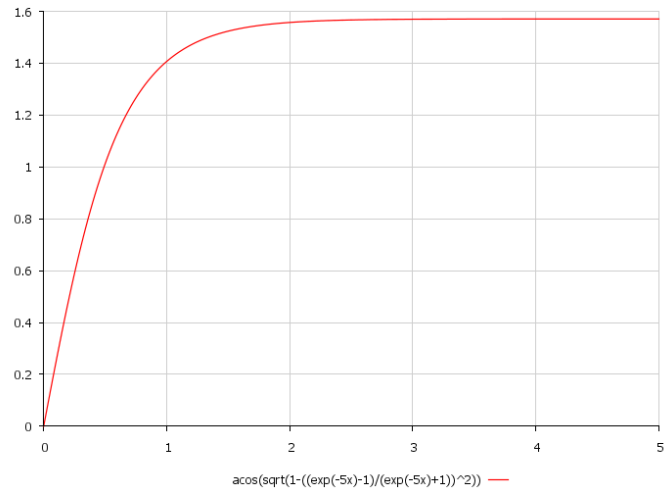

Fig.4. Function θ(*t*): $A = 5$, $\alpha = 1$, see Eqs. (6.1).

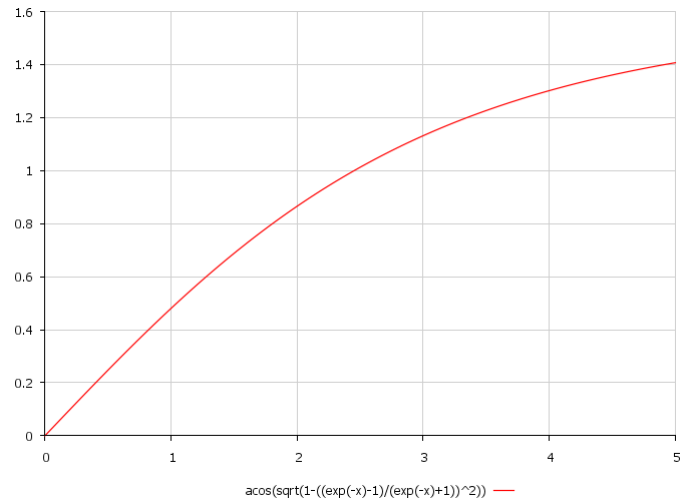

Fig.5. Function θ(*t*): $A = 1$, $\alpha = 1$, see Eqs. (6.1).

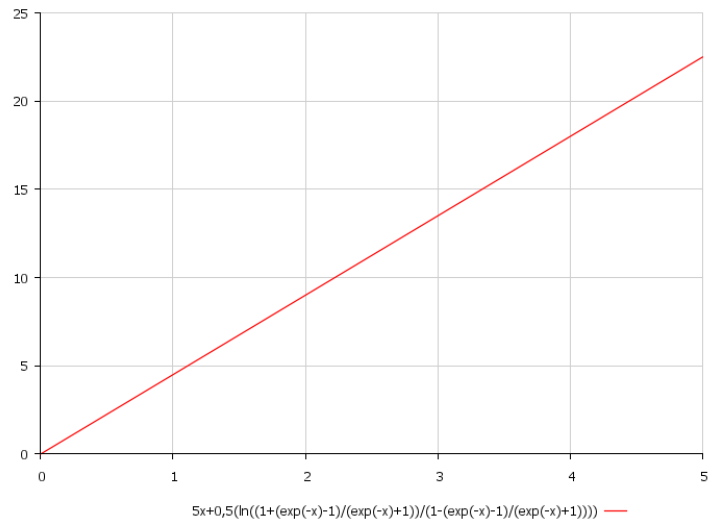

Fig.6. Function ψ(*t*): $(I_1 / I_3) = 5$, $A = 1$, $\alpha = 1$, see Eqs. (6.2).

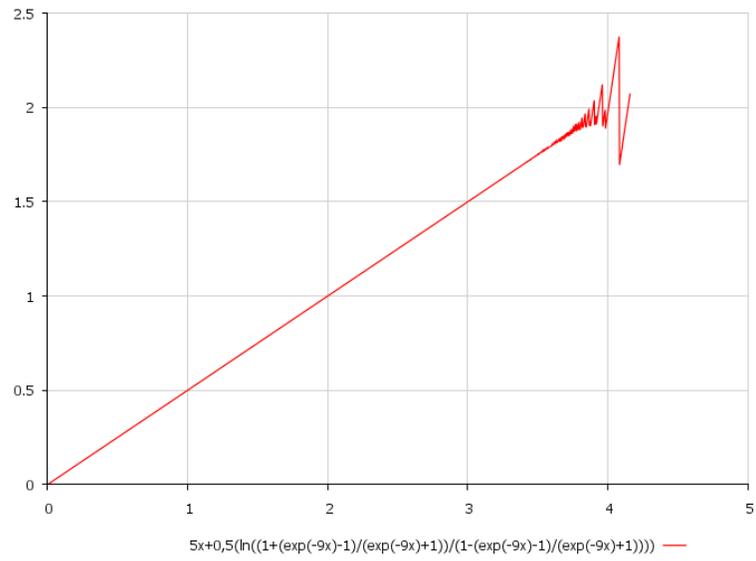

Fig.7. Function $\psi(t)$: $(I_1/I_3) = 5$, $A = 9$, $\alpha = 1$, see Eqs. (6.2).

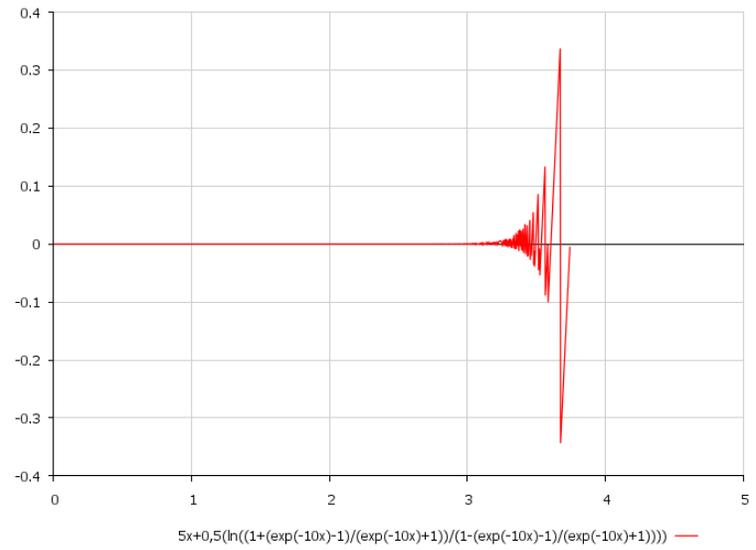

Fig.8. Function $\psi(t)$: $(I_1/I_3) = 5$, $A = 10$, $\alpha = 1$, see Eqs. (6.2).

### 7. <u>Acknowledgements.</u>

Author gratitude to Dr. A.G. Petrov in regard to valuable discussions, which convinced in understanding the fact that it could be important to present even for the specialists (*in rigid body dynamics*) such a non-trivial transformations of the exact solutions from the rotating coordinate system to the fixed Cartesian coordinates, via *Euler's* angles.

For the reason some of them may be having no the sufficient time to execute the calculations properly, it has been made accordingly in the development above.

## **Conflict of interest**